\documentclass[letterpaper,10pt]{article}
\usepackage[margin=1in]{geometry}
\usepackage{float,graphicx,amssymb,amsmath,color,fancybox}
\usepackage[tight,nice]{units}
\usepackage[font={scriptsize,it}]{caption}
\usepackage{wrapfig}
\usepackage{pgfgantt}
\usepackage{tensor}
\usepackage{yfonts}
\usepackage{multicol}
\usepackage{mathrsfs}
\title{Background-Independence from the Perspective\\ of Gauge Theory}
\author{Casey Cartwright\\ Alex Flournoy }

\begin{document}
\maketitle 

\begin{abstract}
We consider two concepts often discussed as significant features of general relativity (particularly when contrasted with the other forces of the Standard Model): background independence and diffeomorphism invariance. We remind the reader of the role of backgrounds both as calculational tools and as part of the formulation of theories. Examining familiar gauge theory constructions, we are able to pinpoint when in the formulation of these theories they become background independent. We then discuss extending the gauge formulation to gravity. In doing so we are able to identify what makes general relativity a background independent theory. We also discuss/dispel suggestions that "active" diffeomorphism invariance is a feature unique to general relativity and we go on to argue against the claim that this symmetry is the origin of background independence of the theory.

\end{abstract}
\begin{multicols}{2}
\section{Introduction}
We seek to clarify what is a surprisingly divisive issue on the role of diffeomorphism invariance in general relativity and in particular how it is related to the background independence of the theory. A cursory survey of both published literature and popular information sources reveals a lack of consensus. The viewpoint of this paper is that by constructing theories based on gauging global symmetries, one can distinguish between the steps of symmetrization and the removal of background independence. This approach is obviously of relevance to the strong and electroweak forces and there is mounting evidence that it applies equally to the gravitational interactions. We approach the discussion at several levels, hoping to make the main points clear to physicists with varying backgrounds (no pun intended). As such, we begin by refamiliarizing the reader with backgrounds used for calculational convenience in the familiar example of Maxwell's electrodynamics using the field equations and in its Lagrangian formulation. The latter allows us to make certain useful definitions. We then turn to the formulation of electrodynamics as a gauge theory where we can most clearly see the separate steps of symmetrization and removal of background independence, and also work in a paradigm immediately applicable to the remaining fundamental forces.     
Our attention then turns to the more subtle case of gravity where we give a brief review of the history and present status of attempts to formulate GR as a gauge theory. This allows us to address claims that diffeomorphism invariance is the source of background independence in the theory.

\section{Backgrounds as calculational tools}
Physicists are perhaps most readily familiar with the notion of backgrounds from calculations in elementary electrodynamics. In that setting a typical calculation involves determining the behavior of a test particle in response to a fixed set of sources. To be more precise, one often posits a (possibly time-dependent) charge distribution to act as a source. This source is used in Maxwell's equations to solve for the relevant field configurations. With the field configurations in hand, the motion of the test charge can be deduced using the Lorentz force law in conjunction with Newton's laws of motion (for nonrelativistic test particles). In such an analysis the backreaction of the sources due to the presence of the test charge is ignored (justified by its having a small mass compared to the sources). In this scheme the source or its induced field configuration serves as a background for the dynamics of the test particle. One can improve the analysis by taking the backreaction into account.

From an action viewpoint this analysis plays out as follows: An action defined with fixed source terms coupled to the undetermined electric and magnetic fields is extremized to obtain equations of motion (Maxwell's) for the fields, which are then solved. These field configurations are then fed into an action for the test particle as fixed background fields, i.e. this action is \textbf{not} extremized with respect to these fields. Extremizing this action with respect to the test particle degrees of freedom then gives equations of motion which (with additional boundary conditions) can be solved to determine the motion of the test particle. Clearly one could forgo the specification of sources and instead specify a set of fixed fields to be used in the test particle action. In either case, the arbitrary assignment of some fixed terms in the test particle action which influence its behavior constitute what can be referred to as a background. This leads us to a working definition of a background: Any degree of freedom appearing in an action with respect to which the action is not extremized constitutes a background. Two comments are in order. First the notion of background here is relevant to the particular action being used. Additionally, this use of the term "background" is synonomous with the term "nondynamical" often encountered in the literature. In this context, taking into account backreaction can be achieved by including all degrees of freedom as dynamical components of a single action. Of course one would still have to stipulate boundary conditions and as such these constitute a sort of background which is only absent in cosmological scenarios \cite{Lee}.

\section{Backgrounds in the formulation of theories}

In the preceding discussion, the split of the action into a background and test particle was a matter of calculational convenience. Our primary concern in this paper is with the presence (and elimination) of backgrounds in the formulation of theories. To keep the discussion more straightforward we will work in the framework of field theory wherein all degrees of freedom are encoded in fields. This approach is also more aligned with our best understanding to date of the electroweak and strong interactions. 

While the gauge symmetry of electromagnetic potentials in relation to electric and magnetic fields is a familiar result (often treated as not much more than a technical detail), its significance in the fundamental structure of electromagnetic interactions is often understated. When one wants to understand electromagnetic interactions and their similarity to the strong and weak interactions of the Standard Model, gauge field theory is the indispensable common setting. Gauge field theory also has the aesthetic advantage of ascribing interactions to the presence of symmetries. Qualitatively the construction of a gauge field theory proceeds as follows: First one specifies a non-interacting field theory which possesses a global (space-time independent) symmetry. The global symmetry is then made local (space-time dependent) typically by modification of the derivative operator to a covariant form with the introduction of a compensating gauge field. The covariant derivative implies a coupling between the original fields and the new gauge field, i.e. an interaction. However at this point the gauge field itself is nondynamical in the sense described above since it has no meaningful variational role in action. To render the gauge field dynamical the action must be augmented to include a gauge kinetic term. With the inclusion of this term, variation of the action with respect to the gauge field becomes a meaningful operation.  

For a specific example, consider the Lagrangian for a free complex scalar field 
\begin{equation}
 L=\frac{1}{2}\partial_{\mu}\phi^*\partial^{\mu}\phi ,
\end{equation}
which enjoys a global phase invariance
\begin{equation}
 \phi\rightarrow e^{i q\alpha}\phi, \hspace{1cm}  \phi^*\rightarrow e^{-i q\alpha}\phi^*.
\end{equation}
Here $q$ is a constant which will eventually play the role of the interaction coupling (or charge) and $\alpha$ is the transformation parameter (like the angle in a rotation). This transformation is only a symmetry of the Lagrangian if the parameter $\alpha$ is independent of spacetime, i.e. if the transformation is "global". Otherwise the derivative generates an additional term due to the change in the parameter, e.g. 
\begin{equation}
 \partial_{\mu}\phi\rightarrow e^{iq\alpha}\partial_{\mu}\phi+iq\partial_{\mu}\alpha e^{iq\alpha}\phi.
\end{equation}
We can promote this global symmetry to a local form with parameters depending on spacetime position by suitably modifying the derivative to a covariant form, i.e. 
\begin{equation}
\nabla_{\mu}\phi'=e^{iq\alpha}\nabla_{\mu}\phi 
\end{equation}
with the addition of a compensating gauge field
\begin{equation}
 \partial_{\mu}\rightarrow \nabla_{\mu}=\partial_{\mu}+iqA_{\mu}.
\end{equation}
Invariance under local transformations of $\phi$ is now guaranteed so long as the gauge field suitably transforms, i.e.
\begin{equation}
 A_{\mu}\rightarrow A'_{\mu}=A_{\mu}+iq\partial_{\mu}\alpha. \label{gaugetrans}
\end{equation}
It is important to take account of the theory to this point. We have a locally invariant Lagrangian which now includes an interaction between the scalar field and the gauge field due to the product of the scalar and gauge fields from the new term in the covariant derivative 
\begin{align}
 \nabla_{\mu}\phi^{*}\nabla^{\mu}\phi=&\partial_{\mu}\phi^*\partial^{\mu}\phi+iq(\phi\partial_{\mu}\phi^{*} - 
\phi^{*}\partial_{\mu}\phi)A^{\mu}\nonumber\\ &+ q^2 A_{\mu}A^{\mu}\phi^{*}\phi.\label{em}
\end{align}
There is however no kinetic (derivative) term for the gauge field, so varying the action with respect to this degree of freedom is meaningless. The gauge field at this point constitutes a background as defined above. To use this action we would have to specify some (arbitrary) gauge field configuration. It is interesting to note that a subset of the choices for the gauge field include those that are gauge equivalent to zero. Such choices may seem to indicate a nontrivial interaction, but actually result from a poor choice of gauge since they are physically indistinguishable from a free scalar field. However these particular choices are also the only ones consistent with the next step in the development of the theory. To proceed, we now provide the gauge field with a locally invariant kinetic term 
\begin{equation}
 L_0=-\frac{1}{4}\tensor{F}{^{\mu}^{\nu}}\tensor{F}{_{\mu}_{\nu}},\hspace{.4cm}\tensor{F}{_{\mu}_{\nu}}=\partial_{\mu}A_{\nu}-\partial_{\nu}A_{\mu}.
\end{equation}
It is now meaningful to vary the action with respect to the gauge field and hence it ceases to be a background for the theory. Referring to the form of the Lagrangian with vanishing kinetic term~(\ref{em}), we can now see that this is a special case of the background independent version, i.e. those instances where the compensating field is gauge equivalent to zero. In some sense this picks out the subset of pure gauge backgrounds as consistent with the fuller formulation of the theory. To this end, it makes sense to start with free Lagrangians in the construction of gauge theories in this manner. This discussion has been purely in terms of the action functional. This story can be given a more geometric underpinning by starting with the bundle construction which identifies the objects from which we build the action as pullbacks of principle gauge and associated vector bundles by local sections. 

An important takeaway from the preceding discussion is the distinction between making a theory locally invariant and rendering it background independent. From the gauge construction this is almost obvious, and indeed this discussion applies point for point to the strong and weak forces in the Standard Model. Nonetheless this has been a source of disagreement when applied to more complicated contexts such as gravitation, to which we now turn.

\section{The case of gravity}

While we can strip away the background dependence of the strong, electromagnetic and weak interactions in the Standard Model, we should be careful to understand the degree to which the results are background independent. After the full gauge construction, the newly introduced compensating fields are removed as backgrounds by rendering them dynamical. However there still occur in the resulting actions degrees of freedom which are pre-selected but not determined by the dynamics of the theory. These include the previously mentioned boundary conditions (which can be addressed by considering "cosmological" theories), but also the geometry of the underlying spacetime as well as its topology and dimension. Every physicist is familiar with the idea that the geometric sector of the remaining backgrounds will be addressed at some level by Einstein's theory of general relativity (GR). But is there any sense in which this part of the story plays out along the familiar lines of gauge theory? Surprisingly, despite a long history and an enormous amount of work, there still remains no complete consensus on how GR is realized as a gauge theory. One aspect of GR that complicates the discussion is that the symmetries expected to be part of the gauge procedure are external, i.e. they act on spacetime, instead of acting on internal degrees of freedom. While there exists reasonably well developed approaches in terms of the action, there are some complications including the unavoidable presence of torsion and the lack of a geometric underpinning in terms of something like a bundle construction. In a forthcoming paper we will address gauge approaches to gravitation (beyond field theory) and what they have to say about the uniqueness of gravitation in contrast to the other Standard Model forces as well as making connections to theories of extended objects. For our arguments here it is sufficient to work at the level of the action in terms of fields.

The history of gauge field theory approaches to gravity actually began shortly after the birth of modern gauge theory, circa the work of Yang-Mills~\cite{yang}. Utiyama first tried to obtain GR by gauging the Lorentz group~\cite{ryoyu}. In his analysis he had to make several unjustified assumptions, but eventually arrived at a theory akin to GR, however energy-momentum was not conserved. Later Sciama gauged the Lorentz group in a theory already containing GR to isolate the role of torsion as reflecting the geometric effects of sources with intrinsic spin~\cite{sciama}. Kibble was the first to fashion the more complete picture by starting in flat spacetime and gauging instead the full Poincare group~\cite{kibble} including not just the Lorentz transformations but spacetime translations as well. His formulation led to the presence of torsion, but also accounted for the more standard elements of GR.  All of these approaches led not to pure GR, but rather to its generalization Einstein-Cartan theory. A decade after Kibble's contribution, Cho fashioned a gauge theory of the translational group $\mathbb{R}^4$ ~\cite{transgauge1}. The resulting theory formulated by Cho is known as the local teleparallel equivalent of general relativity \cite{TGTisnotbetterthenGR}. The success of Kibble's approach has been elaborated upon and generalized in the exhaustive work of Hehl et al(\cite{Hehl2013},\cite{Hehlaffine}). It is interesting to note the duration of development and understanding of external gauge symmetries. Unlike their internal counterparts the merits of external gauge symmetries have been disputed for nearly 50 years. 

At a glance, when one is faced with obtaining GR as a gauge theory, the obvious starting point is flat (non-gravitationally interacting) spacetime and the relevant global symmetry transformations are the Poincare group (the semi-direct product of Lorentz transformations and translations $SO(1,3)\ltimes\mathbb{R}^4$) acting on the spacetime coordinates as well as the Lorentz group acting on the tensor indices of fields. If one tries to mimic too carefully the typical gauge constructions \`{a} la Yang-Mills then one might focus on those transformations that act at a point (linearly), i.e. the Lorentz group. This approach also lends itself more readily to an underlying bundle structure. This was the theme of Utiyama's work. However it was soon pointed out that since the source term in the standard formulation of GR is the conserved energy-momentum tensor, it would be necessary to include the group of translations in the gauge construction. Indeed this argument also reveals why spacetime torsion becomes a necessary ingredient of the final theory anytime the Lorentz group is gauged, since torsion is sourced by spin angular momentum. Technical complications in these gauge constructions include accounting for the action on both tensor indices as well as the coordinates upon which the field configurations depend. This requires working in the tetrad formalism with a local Lorentz frame defined at each point in spacetime. Additionally, since the action itself is integrated over spacetime, its invariance under the local transformations requires not just a modification of the derivative operator, but the promotion of the Lagrangian to a Lagrangian density. In any case, once invariance is achieved, the newly introduced compensating fields are rendered dynamical by adding to the Lagrangian appropriate gauge kinetic terms, e.g. the standard Einstein-Hilbert action. The final technical hurdle of giving the entire program a geometric underpinning in terms of some bundle-like structure is complicated by the translations. Work on this is still ongoing \cite{Ivanenko1983},\cite{Tresguerres2000},\cite{Tresguerres2002},\cite{Sardanashvily2005},\cite{poinc2groupquangrav},\cite{higherTPG},\cite{Tresguerres2014},\cite{DCpro} with several interesting avenues to be discussed more critically in a forthcoming paper (Extended Objects and the Bundle Structure of general relativity; manuscript in preparation). 

What all of these formulations have in common is the gauge field theory approach wherein some global symmetry in a non-interacting theory is gauged resulting in an interacting theory. They also exhibit the key observation above that symmetrizing the theory and making it background independent are two distinct steps. In summary, one can conclude (as many have pointed out already) that the background independence of GR arises from the background geometry being promoted from a fixed input to a true dynamical component of the action. One should keep in mind as mentioned before that the resulting theory still has some residual background dependence in the form of spacetime topology, dimension and any boundary conditions imposed.    

\section{Remarks on the relevance of diffeomorphism invariance}

Rovelli and others have claimed that what makes GR a background independent theory is its invariance under active diffeomorphisms. Here we pose two objections to this conclusion. First, as has been pointed out by numerous authors, the distinction between active and passive diffeomorphisms is ill-conceived. Once one accepts this conclusion, the formulation of gravity as a gauge theory becomes more akin to the standard case. It then follows, from the gauge theory perspective discussed above, that invariance does not itself alleviate backgrounds, but rather making the compensating degrees of freedom dynamical does.

Defining diffeomorphisms as a differentiable maps from one manifold to another (or itself) implies a smooth reassignment of the locations of points in the manifold. The notion of shuffling about the points in spacetime is what some authors call active diffeomorphisms. This is to be (alledgedly) distinguished from starting with one coordinate system and then simply reassigning the coordinates. Rovelli has made a case for the distinction by imagining a sphere (like the surface of Earth) with distinct points labeled by cities. He then imagines a wind map which moves the air over the surface of the Earth to new locations over a period of time. He posits that the wind map is like an active diffeomorphism. From day one to day two the wind could move the gloomy weather from Paris to Denver and the sunny weather from Denver to Paris. On the other hand he then considers coordinatizing the sphere, and in particular assigning Paris and Denver coordinate values, e.g. $(P_1,P_2)$ and $(D_1,D_2)$. At this point he may instead refer to the weather at $(P_1,P_2)$ and $(D_1,D_2)$ without reference to the cities. Now he claims that by choosing new coordinates that swap $(P_1,P_2)\Leftrightarrow (D_1,D_2)$ he has made a change in the weather assignment at these coordinate values, but he clearly has not moved the weather from Paris to Denver and vice versa. This is is Rovelli's basis for distinguishing between active and passive diffeomorphisms.  

The problem with this distinction is the necessity of some underlying unchanging demarcation of points (cities) before the wind map or coordinates are considered. Rovelli himself points out that these cities constitute a background which once removed renders the distinction between active and passive diffeomorphisms needless. The rebuttal to this claim from the perspective of gauge theory is that backgrounds are not unchanging absolutes, but rather nondynamical. To be clear, if we have made an action locally symmetric but not yet given the compensating field a kinetic term, performing a gauge transformation has a nontrivial effect on the compensating field (\ref{gaugetrans}). The field is not absolute, despite being nondynamical! To soften the blow, one may conclude that there are certainly some theories wherein Rovelli's distinction is meaningful, but such scenarios are not part of the any standard gauge theory. To the degree that it seems GR (and indeed the rest of the Standard Model) can be formulated in terms of gauge theory, the need to distinguish between active and passive diffemorphisms is absent.     
\end{multicols}
\clearpage
\setcounter{page}{1}
\bibliography{Gaugebackground}
\bibliographystyle{unsrt}

\end{document}